# RF Transport Electromagnetic Properties of CVD Graphene from DC to 110 MHz

**Shakil Ahmed Awan[1,2], Genhua Pan[1], Laith M. Al Taan[1], Bing Li[1] and Nawfal Jamil[1]**
[1]School of Computing and Mathematics, Faculty of Science and Environment, Plymouth University, Drake Circus, Plymouth, PL4 8AA, UK
[2]Cambridge Graphene Centre, Department of Engineering, University of Cambridge, Madingley Road, Cambridge, CB3 0FA, UK
E-mail: shakil.awan@plymouth.ac.uk

**Abstract**
We report measurement of the radio-frequency (RF) transport electromagnetic properties of chemical vapour deposition (CVD) graphene over the DC to 110 MHz frequency range at room temperature. Graphene on Si/SiO$_2$ substrate was mounted in a shielded four terminal-pair (4TP) adaptor which enabled direct connection to a calibrated precision impedance analyser for measurements. Good agreement is observed for the DC four-probe resistance and the 4TP resistance at 40 Hz, both yielding $R \approx 104$ Ω. In general the apparent graphene channel electromagnetic properties are found to be strongly influenced by the substrate parasitic capacitance and resistance, particularly for high-frequencies $f > 1$ MHz. A phenomenological lumped-parameter equivalent circuit model is presented which matches the frequency response of the graphene 4TP impedance device over approximately seven decades of the frequency range of the applied transport alternating current. Based on this model, it is shown for the first time, that the intrinsic graphene channel resistance of the 4TP device is *frequency-independent* (i.e. dissipationless) with $R_G \approx 105$ Ω or sheet resistance of approximately 182 Ω / □. The parasitic substrate impedance of the device is found shunt $R_G$ with $R_P \approx 2.2$ Ω in series with $C_P \approx 600$ pF. These results suggest that our new RF 4TP method is in good agreement with the conventional DC four-probe method for measuring the intrinsic sheet resistance of single-atom thick materials and could potentially open up new applications in RF electronics, AC quantum Hall effect metrology and sensors based on graphene 4TP devices operating over broad range of frequencies.

## 1. Introduction

Graphene, an atomically thin two-dimensional allotrope of carbon [1], has recently been found to show a number of physical properties, such as high carrier mobility [1-3], ambipolar field-effect [1-5], high current carrying capacity [4] and thermal conductivity [6], which are promising for numerous applications, particularly for nanoelectronics and optoelectronics [7-11]. To date, majority of the effort on nanoelectronics has been focused on DC transport properties of graphene, based on field-effect transistors [12-15], and optical properties of mono and multilayer graphene [16-22]. Although graphene is a promising material for high-frequency applications such as RF transistors [23-26], low-noise amplifiers [27], mixers [28] and frequency doublers [28,29], only a few investigations into the high-frequency transport properties of graphene have been reported [30-33]. Furthermore, given that graphene serves as the channel material in radio-frequency (RF) transistors, significant effort has been devoted to characterization and modeling of RF transistors [24-26, 34-37], rather than characterizing the building block material itself at RF [38,39]. Here, we report on the RF electromagnetic transport properties of monolayer chemical vapour deposition (CVD) graphene carrying transport direct current (DC) and alternating current (AC) with frequencies ranging from 40 Hz to 110 MHz. For some electronic applications of graphene there are three regions of the electromagnetic spectrum which are of key importance for characterising the material properties; DC, RF (covering power frequencies to 300 MHz) and microwaves (covering 0.3-300 GHz). Measurements of graphene at DC are potentially the most precise and relatively easy to implement (such as 2-probe and 4-probe) under variety of applied conditions, such as temperature, magnetic field etc. for elucidating the material or device properties [1-5]. However, DC measurements are limited in providing any information on the dynamic properties of a material or device. In particular, DC measurements can have limited use if the material is being deployed in RF applications where parasitic and contact effects could mask the intrinsic properties preventing detailed understanding of the reactive and dissipative effects in a practical device or system. There are two types of workhorse instruments which enable relatively precise and traceable measurements; impedance analysers (IAs) for RF measurements and vector network analysers (VNAs) for microwave measurements. Although VNAs offer broadband measurements up to at least 110 GHz, their accuracy is limited and depending on the frequency range of operation, it can vary from 1-10% [40]. In contrast, the IAs are limited in frequency range, typically 40 Hz to 110 MHz, but can be precise and accurate (to around 0.1-1% depending on the frequency range) if they are first calibrated using impedance standards of known frequency respons [41-44].

Here, we demonstrate that using a precision four terminal-pair (4TP) impedance analyser with a high-frequency calculable resistance standard (HF-CRS) to calibrate the IA, the intrinsic properties of CVD graphene can be extracted accurately up to 110 MHz for the first time. The results could impact future applications of graphene in electronic devices by enabling optimisation of the production processes, device design, theory, modelling and indeed accurate characterisation. In addition, our results also demonstrate good agreement with the conventional DC four-probe (4P) sheet resistance measurement method [45] for accurately determining the intrinsic resistive properties of graphene in the presence of substrate parasitic impedances. Thus, graphene devices in the 4TP configuration could potentially open up new applications in RF electronics, AC quantum Hall effect metrology and variety of sensors (biological [46], chemical and physical) based on precision impedance spectroscopy [44].

## 2. Design and Fabrication of the Graphene 4TP Impedance Device

Figure 1 shows the design outline and fabricated graphene four terminal-pair (G-4TP) impedance device. Monolayer graphene, confirmed using Raman spectroscopy [47], was placed on a high resistivity (10 kΩ-cm) $Si/SiO_2$ substrate to reduce the effects of substrate parasitic impedances at high- frequencies. However, as it will be shown, in Section 5, the parasitic substrate impedances are found to be finite valued due to the overall size of the device (which enabled connection to the 4TP shielded adaptor). The length and width of the graphene channel is ~ 35 μm between the source-drain or current contacts. Three narrower potential contacts are also patterned on both sides of the graphene channel using photolithography, as shown in Fig. 1. All contacts consist of a 2nm Cr adhesion layer and 80 nm thick Au layer. For a 4TP measurement only potential contacts 1 and 4 as well as the current contacts (labelled source and drain) are used, whilst the other potential contacts (2,3 and 5,6) remained open circuited.

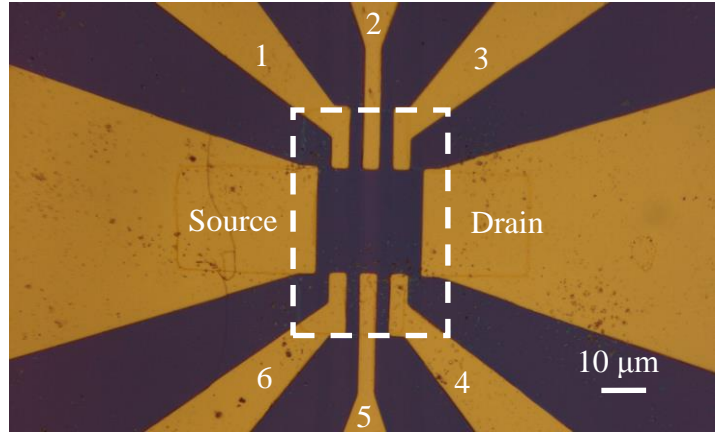

**Figure 1.** Graphene 4TP impedance device on a Si/SiO$_2$ substrate for measurements at frequencies up to 110 MHz. The dashed outline illustrates location of CVD graphene.

## 3. DC Characterisation of the Graphene 4TP Impedance Device

Figure 2 shows the two-probe (2P) and four-probe (4P) DC *IV*-characteristics of the graphene 4TP impedance device. As expected, the $I_{SD}$ versus $V_{SD}$ is linear and the corresponding 2P resistance between the current contacts is found to be $R_{2P} \approx 378\ \Omega$. Similarly, the 4P resistance between the potential contacts (1 and 4), with a channel length of $l = 20$ μm and width $w = 35$ μm, is found to be $R_{4P} \approx 104\ \Omega$, at zero back-gate potential (with carrier density $n \approx 3 \times 10^{12} cm^{-2}$ and mobility $\mu \approx 2000 cm^2 V^{-1} s^{-1}$ obtained using back gate modulation of the $I_{SD}$ current with electric-field [1]). The corresponding sheet resistance (in units of $\Omega / \square$) for a two-dimensional material [45] is given by $R_S = (w/l) \cdot R_{4P}$ and for our device we find $R_S^{DC} \approx 182 \Omega/\square$ at DC and a channel conductivity of $\sigma = (R_S)^{-1} \approx 142 e^2/h$, where $e$ is the electronic charge and $h$ the Planck's constant. After the DC *IV*-characterisation, the G-4TP impedance device was mounted in a shielded adaptor (illustrated in Fig. 3(a)) to enable direct connection with an Agilent 4294A four terminal-pair impedance analyser for measurements up to 110 MHz. The direct connection of the graphene device to the IA is important for two key reasons; firstly this avoids use of coaxial cables to connect the device to IA thereby enabling accurate measurements which would otherwise require cable corrections [44], and secondly the IA is calibrated using a HF-CRS which also makes direct 4TP connections with the IA (further details of this are given in Section 4.0). As the 4TP adaptor requires high and low current contacts and two high and low potential contacts (1 and 4, respectively), the remaining potential contacts were left open circuited during RF measurements. At microwave frequencies these open circuit leads may give rise to resonances and parasitic effects, however at the lower ~100 MHz frequencies, these are found to have negligible effect on the measured transport properties of our graphene 4TP impedance device. The definition of a 4TP impedance was first devised by Cutkosky in 1964 [48]. It is possible to rewrite the 4TP definition for a general linear network as shown in Fig. 3(*a*), such that:

$$\begin{bmatrix} V_1 \\ V_2 \\ V_3 \\ V_4 \end{bmatrix} = \begin{bmatrix} Z_{11} & Z_{12} & Z_{13} & Z_{14} \\ Z_{21} & Z_{22} & Z_{23} & Z_{24} \\ Z_{31} & Z_{32} & Z_{33} & Z_{34} \\ Z_{41} & Z_{42} & Z_{43} & Z_{44} \end{bmatrix} \cdot \begin{bmatrix} I_1 \\ I_2 \\ I_3 \\ I_4 \end{bmatrix} \qquad (1)$$

where *V* and *I* are the voltages and currents at the four respective ports and $Z_{ij} \equiv V_i/I_j$, so that $I_{j \neq k} = 0$ with $1 \leq (i, j, k) \leq 4$, such that for a linear reciprocal 4TP network $Z_{ij} = Z_{ji}$ are the two terminal-pair (2TP) impedance parameters. Thus, the four terminal-pair impedance is:

$$Z_{4TP} = \left. \frac{V_2}{I_4} \right|_{I_2 = I_3 = V_3 = 0} \qquad (2)$$

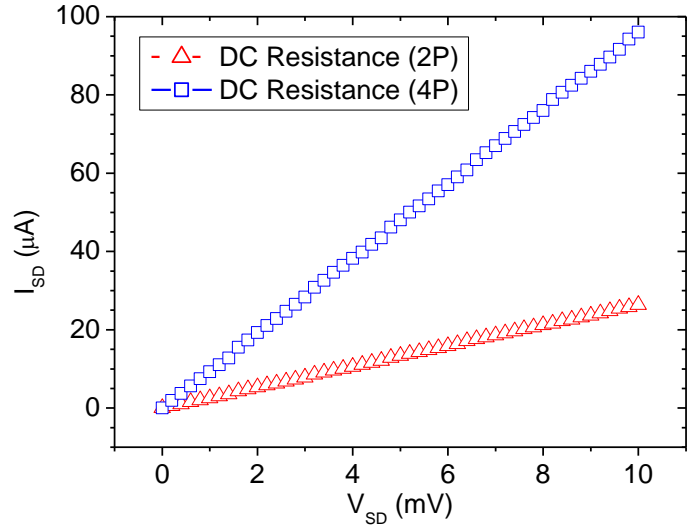

**Figure 2.** DC two-probe (source-drain)) and four-probe (using potential contacts 1 and 4) *IV*-characteristics of the graphene 4TP impedance device. The 2P resistance is found to be $\approx 378\ \Omega$ whereas the 4P resistance is found to be $\approx 104\ \Omega$, at zero back-gate potential.

Also shown in Fig. 3(b) is the RF equivalent circuit phenomenological lumped-parameter model of the graphene 4TP impedance device. The significance and suitability of this circuit model for the G-4TP impedance device is discussed in Section 5.0.

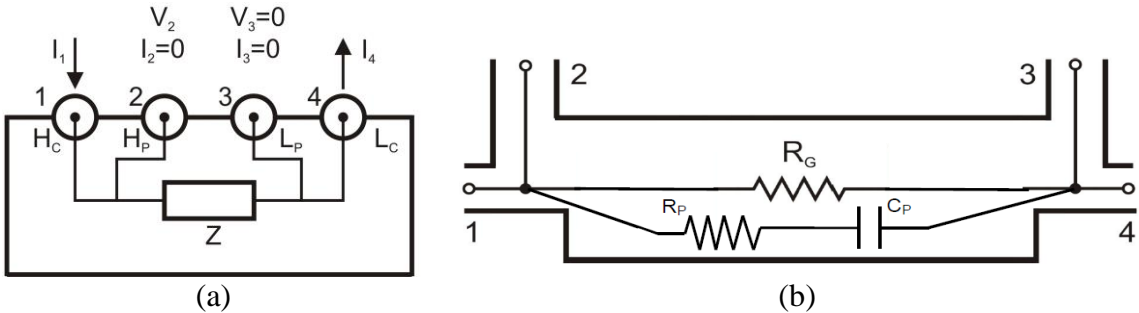

(a)          (b)

**Figure 3.** Four terminal-pair definition of impedance *Z* or device-under-test mounted in a shielded adaptor (a) and the RF equivalent circuit phenomenological lumped-parameter model of the graphene 4TP impedance device consisting of channel resistance $R_G$ with substrate parasitic impedances $R_P$ and $C_P$ (b). Ports labelled 1-4 in both diagrams are often also referred to as high-current ($H_C$), high-potential ($H_P$), low-potential ($L_P$) and low-current ($L_C$), respectively.

of standards were specifically developed to calibrate and to verify the correct operation of IAs and LCR meters from DC to 110 MHz (and to enable robust comparison of such instruments from different commercial suppliers). The frequency response of the HF-CRS can be calculated from first principles based on Maxwell's equations and the full analytical derivation of the expressions for frequency-dependent resistance, inductance and capacitance are given in [44]. The 4TP impedance of the standard is given by:

$$Z_{4TP}(\omega) = \frac{R + i\omega\left[L(1-\omega^2 LC) - CR^2\right]}{1+\omega^2\left[C^2R^2 - 2LC + \omega^2 L^2 C^2\right]} \quad (3)$$

where $R$ is the measured DC resistance of the standard, $i$ is the unit complex number, $\omega = 2\pi f$, inductance $L = 0.19$ µH and capacitance $C = 2.1$ pF are the calculated values of the standard based on the dimensions and geometry of the standard. Figure 5 shows the measured apparent resistance and reactance of the 100 Ω HF-CRS, using Agilent 4294A, up to 110 MHz. Good agreement is observed between the measured and calculated frequency response of the HF-CRS, for both resistance and reactance, up to 110 MHz, confirming accurate calibration of the range resistors (to ~1% for $f < 100$ MHz, and ~5% for $f > 100$ MHz). An instrument artefact occurs at 15 MHz, as shown in the measured apparent resistance and reactance data. This is due to the instrument switching between its two internal bridge measurement systems (which the manufacturer refers to as auto-balancing and I-V bridges) covering the frequency ranges above and below 15 MHz, respectively.

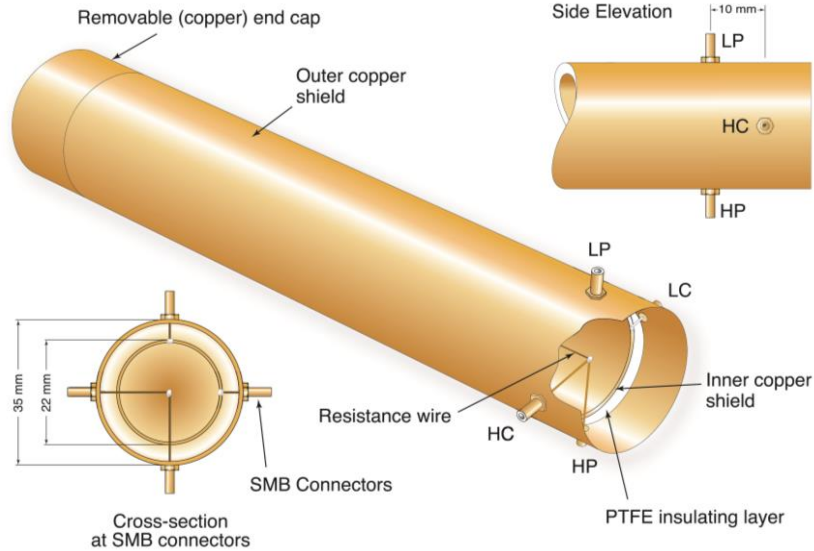

**Figure 4**. Schematic of the coaxial 4TP high-frequency calculable resistance standard (with permission from *IEEE* [41]).

## 5. Characterisation of the CVD Graphene 4TP Impedance Device

Figure 6 shows the measured and modelled frequency response of the graphene 4TP impedance device from 40 Hz to 110 MHz (with zero back-gate potential) at room temperature. At 40 Hz the measured 4TP resistance of ≈ 105 Ω is in good agreement with the 4P DC measured resistance of ≈ 104 Ω. For the entire measurement frequency range, the apparent resistance and reactance show good agreement with the model of Fig. 3(b), using least-squares fitting parameters $R_G = 105$ Ω, $R_P = 2.2$ Ω and $C_P = 600$ pF (from the device size and geometry a simple calculation gives $C_P \approx 530$ pF). The measurement results shown in Fig. 6 and the equivalent circuit model of Fig. 3(b) suggest that any inductive or capacitive components for the graphene channel are negligible for this device of channel length $l = 20$ μm and upper operational frequency of 110 MHz. It is worth noting that as the frequency increases above $f \sim 1$MHz the apparent resistance tends to $\sim 2$ Ω (since the parasitic capacitance begins to short the graphene channel resistance) while the apparent reactance tends to zero because the short graphene channel has negligible capacitive or inductive impedance (within our experimental resolution). The relatively small differences between the measured apparent resistance and reactance with the simulated values, based on the model of Fig. 3(b), may be due to the possibility that the simple model of Fig. 3(b) may need further optimisation in future work. The frequency-dependent response of G-4TP in Fig. 6, for both apparent resistance and reactance, demonstrates that the substrate parasitic capacitance and resistance have a strong influence on the apparent electromagnetic behaviour of the graphene channel, particularly at higher-frequencies ($f > 1$MHz). However, comparison of the observed apparent resistance and reactance of the graphene 4TP device with the model of Fig. 3(b) demonstrate that the intrinsic channel resistance of graphene is *frequency-independent* such that $R_G \approx 105$ Ω, approximately the same as the DC four-probe value, giving a sheet resistance of $R_S^{RF} = R_S^{DC} \approx 182 \Omega/\square$. This result suggests that our new RF 4TP method is in good agreement with the conventional DC 4-probe method for measuring sheet resistance of single-atom thick materials such as graphene, with the added benefit of enabling accurate and robust measurements of the dynamic and practical material, device and system-level properties, particularly for RF electronics, AC quantum Hall effect and biosensor applications. In addition, the frequency-independent resistance of CVD graphene reported here is also in good agreement with our measurements of exfoliated graphene samples integrated into coplanar waveguides (calibrated against National Institute of Standards and Technology (NIST) traceable SOLT standards) operating over the DC to 13.5 GHz frequency range, which also show frequency-independent channel resistance [38]. An interesting possibility emerges based on these findings with regards long sought after quantum standard of impedance (QSI) [41,44]. To date it has been extremely challenging to develop a QSI based on conventional GaAs/GaAlAs devices, see for example [49], due to the observed finite $10^{-7}$/kHz frequency coefficient arising from dissipative capacitive charging currents [50]. However, our measurements both up to 110 MHz and 13.5 GHz demonstrate that the graphene resistance is intrinsically frequency-independent and is ideally suited to the metrological QSI applications. Thus, further work is needed on a variety of other geometries, configurations and applied conditions (such as field, current, temperature etc.) of graphene 4TP impedance devices, as well as exfoliated graphene, in addition to the results reported here for CVD graphene. This would be useful towards developing a comprehensive model of the intrinsic RF electromagnetic properties of graphene, enable robust comparisons with theoretical models (based on Boltzmann transport theory), and therefore may contribute to numerous practical applications of graphene in the near future.

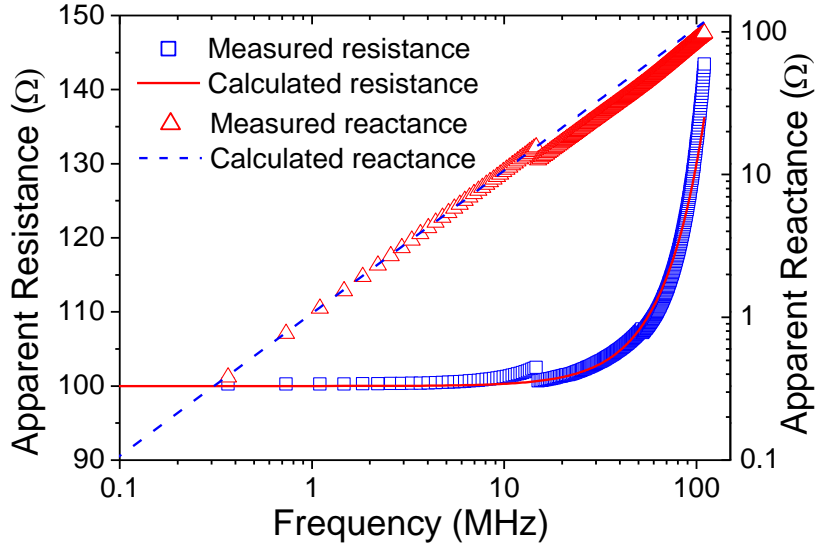

**Figure 5.** Measurement of the 4TP HF-CRS on Agilent 4294A at frequencies up to 110 MHz. The solid and dashed lines are the calculated response of the HF-CRS based on equation (3) [44].

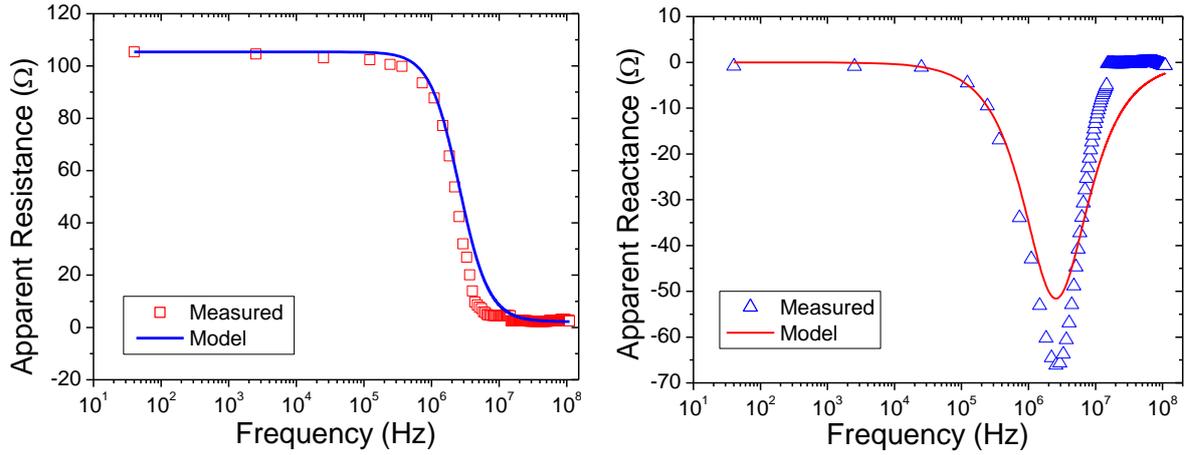

**Figure 6**. Measured and modelled apparent resistance and reactance of the graphene 4TP impedance device from 40 Hz to 110 MHz. The solid line is the model of Fig. 3(b) with fitting parameters $R_G \approx 105\ \Omega$, $R_P \approx 2.2\ \Omega$ and $C_P \approx 600$ pF.

## 6. Conclusion

The RF transport electromagnetic properties of CVD graphene have been measured from DC to 110 MHz at room temperature using a calibrated precision 4TP impedance analyser and a calculable high-frequency reference resistance standard. The 4TP low-frequency AC and four-terminal DC resistance of graphene are found to be approximately the same within 1 %. In general the response of the graphene 4TP impedance device was found to be strongly dependent on the substrate parasitic impedance. A lumped parameter equivalent circuit model was used to fit the measured data for apparent resistance and reactance which yielded an *intrinsic*, frequency-independent (i.e. dissipationless compared to Joule losses at DC), graphene channel resistance of $R_G \approx 105\ \Omega$ or sheet resistance of $R_S^{RF} \approx 182\ \Omega/\square$ shunted by substrate parasitic impedance of $R_P \approx 2.2\ \Omega$ in series with $C_P \approx 600$ pF. The RF 4TP sheet resistance value is found to be in good agreement with the measured DC four-probe method, i.e. $R_S^{RF} = R_S^{DC}$ over seven decades of the frequency of the applied transport alternating current. This suggests our new RF 4TP method may be applied more broadly for the investigation of the dynamic and practical graphene material, device and system level properties for range of potential applications in RF electronics, AC quantum Hall effect metrology and sensors.

## 7. Acknowledgements
The authors acknowledge Dr Bryan P. Kibble, Dr David Hasko, Dr Tawfique Hasan, Dr Antonio Lombardo and Prof Mohammed Z. Ahmed for useful discussions and Graphene Square Inc. for providing the graphene sample.